\documentclass [10pt, twocolumn, conference] {IEEEtran}
\usepackage{amssymb}
\usepackage{graphicx}
\usepackage{multirow}
\usepackage{epstopdf}
\usepackage{setspace}
\usepackage{booktabs,caption,fixltx2e}
\usepackage[flushleft]{threeparttable}
\usepackage{algorithm,algpseudocode}
\usepackage{enumitem}
\usepackage{amsmath}
\usepackage{comment}
\usepackage{subcaption}
\usepackage{caption}
\usepackage{balance}
\usepackage{siunitx}

\usepackage{diagbox}
\graphicspath {{./images}}
\algrenewcomment[1]{\bgroup\hfill//~#1\egroup}

\usepackage{color}
\usepackage{romannum}
\usepackage{color,soul}
\usepackage{algorithm}
\usepackage{algpseudocode}
\usepackage[braket, qm]{qcircuit}

\begin{document}

\title{Enhancing Quantum Dense Coding Robustness Using Information Entropy-Based Metrics}

\author{Syed Emad Uddin Shubha, Tasnuva Farheen\\Division of Computer Science, Louisiana State University}

\maketitle

\begin{abstract}

 Superdense Coding is a cornerstone in secure quantum communication, exploiting pre-shared entanglement to encode two classical bits within a single qubit. However, noise and decoherence deteriorate entanglement quality, restricting both fidelity and channel capacity in practical settings. Traditional methods, such as error correcting codes or entanglement distillation, are generally inadequate for dynamically varying noise conditions. Moreover, reliance on fidelity alone may fail to capture more subtle noise effects. This work introduces an adaptive protocol that integrates the five-qubit perfect code with a novel global adaptive purification that avoids discarding entangled pairs. By monitoring two information entropy-based metrics, quantum discord (QD) and entanglement of formation (EoF) from pilot pairs, we dynamically tune a global unitary to counteract noise. Our simulations, under both amplitude and phase damping, indicate that this integrated strategy could significantly enhance superdense coding robustness while preserving high throughput, thereby offering a scalable pathway toward a high-capacity quantum internet.

\end{abstract}


\vspace{-1em}

\section{Introduction}

Classical communication typically transmits one bit per binary signal, though advanced modulation can boost capacity. Superdense coding \cite{bennett1992communication}, by contrast, exploits quantum entanglement to send two classical bits with a single qubit \cite{bowen2001classical}. This efficiency reduces the required physical resources, making superdense coding particularly useful for quantum networking, satellite communications, and secure data transfer. This protocol has been experimentally demonstrated using photon polarization entanglement \cite{mattle1996dense} and further adapted for practical applications over optical fiber links. However, noise significantly impacts channel capacity and fidelity, as observed in experimental attempts, often reducing performance to below the theoretical maximum, such as roughly 80\% in early demonstrations \cite{shadman2012distributed, barreiro2008beating, williams2017superdense}.

Traditional approaches to combat noise include quantum error correction~(QEC) and entanglement purification. Although QEC can protect a qubit from single-qubit errors, it struggles with multi-qubit errors \cite{dur2007entanglement}. 
On the other hand, conventional purification schemes~\cite{bennett1996purification, deutsch1996quantum} have limited yield due to their reliance on local operations and classical communication (LOCC). They distill multiple noisy Bell pairs to construct a high-fidelity one. Although it is useful for teleportation but applying in superdense coding significantly reduces the channel capacity.

 Recently, adaptive quantum error correction (QEC) protocols have been proposed to dynamically mitigate noise by modeling its behavior and adjusting correction codes accordingly \cite{ghosh2018automated, wang2022automated}. Building on these successes, similar ideas have been explored in the context of dense coding \cite{sadlier2016superdense, nilesh2022automated}. In this work, we extend the adaptive concept to purification. We combine a five-qubit code with a global purification step that adapts its parameters in real time. The key question, then, is how to make purification “adaptive” without resorting to discarding pairs.

To address this, we use \emph{quantum discord} (QD) and \emph{entanglement of formation} (EoF) as real-time metrics for noise. QD has previously been used in quantum key distribution \cite{pirandola2014quantum} and EoF has been studied in teleportation \cite{plenio1998teleportation}. Those literature have motivated us to study their role in
superdense coding, and we found a strong correlation of these metrics with fidelity. This finding suggests that they can likewise guide purification in dense coding. By monitoring QD and EoF, our protocol detects the severity of noise damping, then tunes the purification parameters accordingly, thereby retaining more entangled pairs and maintaining higher channel capacity than conventional fixed-parameter approaches.

Our protocol introduces "pilot pairs" which we define as additional entangled qubits that do not carry data but undergo the same quantum evolution as the main qubits. This setup enables Bob to measure QD and EoF, thereby gauging the noise pattern. Bob then applies the adaptive purification to coherently reverse the majority of noise effects without sacrificing data carrying pairs. The perfect code continues to protect against single qubit errors, while the global purification addresses residual multi-qubit noise.

\noindent\textbf{Contributions.} We summarize main contributions below:
\begin{enumerate}
[leftmargin=*]
    \item We propose a novel protocol integrating QEC with adaptive purification. The adaptive purification uses quantum correlations as metrics to dynamically characterize noise.
    \item We have shown quantum correlations, like QD and EoF are strongly correlated with fidelity and, thus, can work as noise indicators to tune the adaptive purification.
    \item Our adaptive purification protocol aims to correct errors without distillation of entanglement pairs, taking advantage of non-local operations.
\end{enumerate}

The remainder of this paper is organized as follows. First, we provide a concise theoretical description of all the key terms. Next, we describe our integrated protocol and methodology in detail. Finally, we present mathematical results that validate our method and outline future directions for scaling this adaptive architecture in more complex quantum networks.

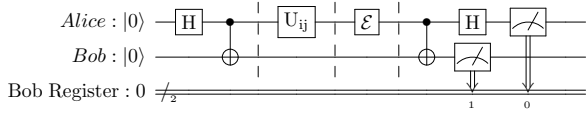
\begin{figure}[htb]
    \centering
\scalebox{0.75}{
    \Qcircuit @C=1.0em @R=0.2em @!R { \\
        \nghost{{Alice} : \ket{{0}} } & \lstick{{Alice} : \ket{{0}} } & \gate{\mathrm{H}} & \ctrl{1} \barrier[0em]{1} & \qw & \gate{\mathrm{U_{ij}}} \barrier[0em]{1} & \qw & \gate{\mathcal{E}} \barrier[0em]{1} & \qw & \ctrl{1} & \gate{\mathrm{H}} & \meter & \qw & \qw\\
        \nghost{{Bob} : \ket{{0}} } & \lstick{{Bob} : \ket{{0}} } & \qw & \targ & \qw & \qw & \qw & \qw & \qw & \targ & \meter & \qw & \qw & \qw\\
        \nghost{\mathrm{{Bob\;Register} : 0 }} & \lstick{\mathrm{{Bob\;Register} : 0 }} & \lstick{/_{_{2}}} \cw & \cw & \cw & \cw & \cw & \cw & \cw & \cw & \dstick{_{_{\hspace{0.0em}1}}} \cw \ar @{<=} [-1,0] & \dstick{_{_{\hspace{0.0em}0}}} \cw \ar @{<=} [-2,0] & \cw & \cw\\
    \\
    }}
    \vspace{-1em}
    \caption{Superdense Coding scheme. If $\mathcal{E}= I$, Bob will receive correct bitstring $ij$. However, for noisy channels, $\mathcal{E} \neq I$.}
    \label{fig:Superdense}
\end{figure}

\section{Background and Technical Overview}
In this section we provide a theoretical overview of the superdense protocol and related concepts. We discuss the noise channels and quantum correlations used in this work. Since Quantum Error Correction and Adaptive Purification are core part of our architecture, we have discussed the existing literature as well.

\subsection{Superdense Coding}
Superdense coding is a quantum communication protocol that enables the transmission of two classical bits by sending a single qubit ~\cite{bowen2001classical}, provided that a maximally entangled state is shared between the sender and the receiver. We assume two parties share a Bell state:
\begin{equation}
    |\Phi^+\rangle = \frac{1}{\sqrt{2}} \left( |00\rangle + |11\rangle \right)
\end{equation}

The sender (Alice) encodes her two-bit classical message $ij$ by applying unitary $U_{ij}$ to her qubit. The encoded qubit is then transmitted to the receiver (Bob), who performs a Bell basis measurement on the two qubits to retrieve the classical message. Here, 
\begin{equation}
    U_{ij} = Z^i X^j, \quad i,j \in \{ 0,1 \}
\end{equation}

Fig.~\ref{fig:Superdense} shows the circuit for the superdense coding. If $\mathcal{E} = I$ (no error), then Bob receives exactly the same classical bit $ij$. However, usually that doesn't happen due to noise.

The channel capacity \cite{bruss2004distributed} of a superdense coding  protocol can be expressed as:
\vspace{-0.75em}
\begin{equation}
    C = \log_2 d_A + S(\rho_B) - S(\rho_{AB})
\end{equation}
where $d_A$ is the dimension of Alice's subsystem, $\rho_B$ is the reduced density matrix of Bob's subsystem, and $\rho_{AB}$ is the joint state of the system. For a maximally entangled Bell state, we have $d_A = 2$, $S(\rho_B) = 1$, and $S(\rho_{AB}) = 0$, leading to $ (\log_2 2 + 1 - 0 )=$ \textbf{2 bits}.

\subsection{Quantum Noise Channels}
Quantum noise channels are described by completely positive, trace-preserving (CPTP) maps \cite{nielsen2010quantum}. Such channels can be expressed in the Kraus representation:
\begin{equation}
    \mathcal{E}(\rho) = \sum_{i} K_i \rho K_i^\dagger, \quad \text{with} \quad \sum_i K_i^\dagger K_i = I.
\end{equation}
A quantum channel can have two different Kraus representations; however, the Choi representations are the same.

\subsubsection*{Amplitude Damping Channel:}
The amplitude damping channel models energy dissipation (e.g., spontaneous emission \cite{chessa2021quantum}). Its Kraus operators are given by:
\begin{equation}
    A_0 = \begin{pmatrix} 1 & 0 \\ 0 & \sqrt{1-p} \end{pmatrix}, \quad
    A_1 = \begin{pmatrix} 0 & \sqrt{p} \\ 0 & 0 \end{pmatrix}
\end{equation}
where $p$ ($0 \le p\le 1$) represents the probability of energy loss. 

\subsubsection*{Phase Damping Channel:}

The phase damping channel describes the loss of quantum coherence \cite{dutta2023qudit} without energy dissipation. Its Kraus operators are:
\begin{equation}
    P_0 = \sqrt{1-q}\, I, \quad
    P_1 = \sqrt{q}\, \begin{pmatrix} 1 & 0 \\ 0 & 0 \end{pmatrix}, \quad
    P_2 = \sqrt{q}\, \begin{pmatrix} 0 & 0 \\ 0 & 1 \end{pmatrix}
\end{equation}
where $q$ ($0 \le q \le 1$) is the probability of phase damping. 

These noise models provide a framework for analyzing the effects of realistic quantum channels. At last, we need a metric to determine how close two quantum states (i.e., $\rho$, $\sigma$) are \cite{dur2007entanglement}. We can define this using Fidelity:
\begin{equation}
    F(\rho, \sigma) = \left( Tr \sqrt{\sqrt\rho \sigma \sqrt \rho} \right)^2
\end{equation}

\subsection{Quantum Correlations}
Quantum correlations are the statistical relationships between quantum measurements that classical physics cannot explain. Correlations like
Quantum Discord and Entanglement of Formations can reveal many properties of quantum states \cite{cen2011quantifying, horodecki2009quantum} such as non-locality and non-classicality.

\subsubsection*{Entanglement of formation:} 
Given a bipartite state $\rho$, Entanglement of formation (EoF) quantifies the minimum entanglement needed to prepare it from pure entangled states \cite{plenio1998teleportation}. We can define EoF as:
\begin{equation}
    E_F(\rho) = \text{min} \sum_i p_iE(\ket{\psi_i})
\end{equation}
where the minimum is taken over all possible decompositions of $\rho = \sum_i p_i \ket{\psi_i}\bra{\psi_i}$ and $E(\ket{\psi})$ is the Entanglement entropy of the pure state $\psi$.

EoF is zero for separable states and increases with entanglement strength. For pure states, it equals the von Neumann entropy of one subsystem.

\subsubsection*{Quantum Discord:}
In bipartite quantum systems, Quantum Discord (QD) measures quantum correlations beyond entanglement. QD is defined as the difference between quantum and classical mutual information \cite{cen2011quantifying}:
\begin{equation}
    \delta(A:B) = I(A:B) - J(A:B),
\end{equation}
Here $I(A:B)$ is the quantum mutual information between subsystem $A$ and $B$. And $J(A:B)$ is the classical mutual information, defined as the maximum classical correlation obtainable via local measurements. $\delta(A:B) \neq \delta(B:A)$ since it depends on the subsystem being measured. Discord can be non-zero even for separable states, making it a useful resource in quantum communication.

\subsection{Quantum Error Correction}
Quantum error correction (QEC) is essential for protecting quantum information from decoherence and noise. In the stabilizer formalism, a quantum code is defined as the common $+1$ eigenspace of an abelian subgroup $S$ of the $n$-qubit Pauli group $\mathcal{P}_n$. Logical qubits are encoded in a subspace (called coding space), and errors are detected by measuring the stabilizer generators \cite{nielsen2010quantum, roffe2019quantum} without disturbing the encoded information. Say $S=\langle s_i \rangle$, where $s_k$'s are the stabilizer generators and say $\{ E_j\}$ are the set of error the code can correct. Then, for a coding space $\mathcal{H}_c$, we have:
\begin{equation}
     \forall |\psi\rangle \in \mathcal{H}_c \implies  s_i E_j |\psi\rangle= c_{ij} E_j |\psi\rangle
\end{equation}
 $\{c_{ij}\}$ are called syndrome for a specific $j$, with $c_{ij} \in \{+1,-1\}$ and $c_{ij}=+1$ for all $i$ when $E_j = I$.

Say $|0\rangle_L$ and $|1\rangle_L$ are logical qubits with logical bit flip $X_L$ and logical phase flip $Z_L$. They are defined as follows:
\vspace{-1.5em}
\begin{center}
    
\begin{equation}
\begin{split}
\mathcal{H}_c = span \left(|0\rangle_L , |1\rangle_L  \right) \\
    Z_L |0\rangle_L = |0\rangle_L, Z_L |1\rangle_L = -|1\rangle_L \\
    X_L |0\rangle_L = |1\rangle_L, X_L |1\rangle_L = |0\rangle_L
\end{split}
\end{equation}
\end{center}

A notable example is the \emph{five-qubit code}, the smallest code capable of correcting any arbitrary single-qubit errors \cite{knill2001benchmarking}. This code encodes one logical qubit into five physical qubits and is called \emph{perfect} because it saturates the quantum hamming bound. Its stabilizer generators are given by Eq.~\ref{Eq:Stab5}.
\vspace{-0.5em}
\begin{equation}
\begin{split}
        g_1 &= X \, Z \, Z \, X \, I, \\
    g_2 &= I \, X \, Z \, Z \, X, \\
    g_3 &= X \, I \, X \, Z \, Z, \\
    g_4 &= Z \, X \, I \, X \, Z.
\end{split}
\label{Eq:Stab5}
\end{equation}

We can define the logical erros as $X_L = XXXXX, Z_L = ZZZZZ$. It can be proven that if an error correcting code can correct both Bit Flip and Phase Flip Error then it can correct all single qubit error. Hence, for the perfect code, each generator $g_i$ commutes with each others, and any single-qubit error anti-commutes with at least one of these generators, ensuring that the error can be uniquely identified and corrected. 

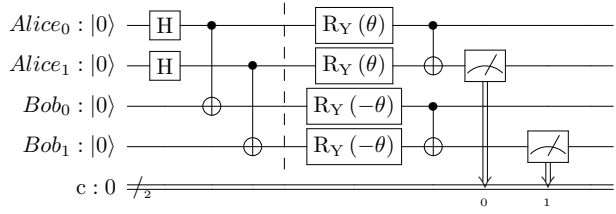
\begin{figure}[tbp]
    \centering
    \scalebox{0.85}{
    \Qcircuit @C=1.0em @R=0.2em @!R { \\
	 	\nghost{{Alice}_{0} : \ket{{0}} } & \lstick{{Alice}_{0} : \ket{{0}} } & \gate{\mathrm{H}} & \ctrl{2} & \qw \barrier[0em]{3} & \qw & \gate{\mathrm{R_Y}\,(\mathrm{{\ensuremath{\theta}}})} & \ctrl{1} & \qw & \qw & \qw & \qw\\
	 	\nghost{{Alice}_{1} : \ket{{0}} } & \lstick{{Alice}_{1} : \ket{{0}} } & \gate{\mathrm{H}} & \qw & \ctrl{2} & \qw & \gate{\mathrm{R_Y}\,(\mathrm{{\ensuremath{\theta}}})} & \targ & \meter & \qw & \qw & \qw\\
	 	\nghost{{Bob}_{0} : \ket{{0}} } & \lstick{{Bob}_{0} : \ket{{0}} } & \qw & \targ & \qw & \qw & \gate{\mathrm{R_Y}\,(\mathrm{{\ensuremath{-\theta}}})} & \ctrl{1} & \qw & \qw & \qw & \qw\\
	 	\nghost{{Bob}_{1} : \ket{{0}} } & \lstick{{Bob}_{1} : \ket{{0}} } & \qw & \qw & \targ & \qw & \gate{\mathrm{R_Y}\,(\mathrm{{\ensuremath{-\theta}}})} & \targ & \qw & \meter & \qw & \qw\\
	 	\nghost{\mathrm{{c} : 0 }} & \lstick{\mathrm{{c} : 0 }} & \lstick{/_{_{2}}} \cw & \cw & \cw & \cw & \cw & \cw & \dstick{_{_{\hspace{0.0em}0}}} \cw \ar @{<=} [-3,0] & \dstick{_{_{\hspace{0.0em}1}}} \cw \ar @{<=} [-1,0] & \cw & \cw\\
    \\
    }}
    \vspace{-1em}
    \caption{DEJMPS Protocol where $\theta$ is usually set as $\pi/2$.}
    \label{fig:DEJMPS}
\end{figure}

\subsection{Entanglement Distillation}
Entanglement distillation (also known as purification) is the process of converting multiple copies of a noisy entangled state into a smaller number of states with higher fidelity \cite{bennett1996purification, horodecki2009quantum},  using only local operations and classical communication~(LOCC). 
The BBPSSW protocol was one of the first entanglement purification schemes ~\cite{bennett1996purification} and uses bilateral CNOT operations and measurements. In this protocol, two copies of $\rho_{AB}$ are combined; by comparing measurement outcomes on selected qubits, the protocol probabilistically discards pairs with low fidelity and retains those with higher fidelity. An improvement over BBPSSW is offered by the DEJMPS protocol, as shown in Fig.~\ref{fig:DEJMPS}, which includes additional local unitary operations prior to CNOT gates ~\cite{deutsch1996quantum}, which are particularly effective at correcting phase errors. This adaptation leads to an increased yield of high-fidelity pairs, making the protocol more robust \cite{dur2007entanglement} in the presence of realistic noise. Usually they apply $R_Y(\theta)$ with $\theta=\frac{\pi}{2}$. If someone starts with Fidelity  $F>0.5$, then in the next round the fidelity becomes ($p_s$ is the probability of success):
\begin{equation}
    \begin{split}
        F'= \frac{F^2 + (1-F)^2/9}{p_s} \\
        p_s = F^2 + \frac{2}{3} F(1-F) + \frac{5}{9} (1-F)^2
    \end{split}
\end{equation}

\vspace{-1em}

\section{Methodology}
\subsection{Quantum Communication Channel Modeling} In this work, we model a realistic quantum communication channel incorporating two primary noise processes: amplitude damping (AD) and phase damping (PD), characterized by error parameters $p$ and $q$, respectively. The amplitude damping channel, which accounts for energy dissipation (e.g., spontaneous emission), is given by $AD$.
Similarly, the phase damping channel, responsible for the loss of quantum coherence without energy loss, is represented by $PD$.
To capture the combined effect of these noise processes, we define a composite noise channel that symmetrically integrates both mechanisms. Since the ordering is equally likely, given an input state $\rho$, the composite channel is taken as
\begin{equation}
    \mathcal{E}(\rho) = \frac{1}{2}\left( AD(PD(\rho)) + PD(AD(\rho)) \right) \otimes \mathcal{I}
\end{equation}

We apply this composite channel to an initially prepared maximally entangled Bell state $|\Phi^+\rangle$, thereby simulating the realistic degradation of entanglement in a quantum channel.

\vspace{-0.78em}

\subsection{Entropy Metrics Calculation} To quantify entanglement and correlations, we calculate two metrics: \emph{Entanglement of Formation}~(EoF) and \emph{Quantum Discord}~(QD).
The EoF measures the amount of entanglement in a quantum state, while the QD captures the total quantum correlations, including those beyond entanglement.

\textbf{Quantum Discord:} To calculate QD, defined as \(\delta(A:B) = I(A:B) - J(A:B)\), we need both quantum and classical mutual information. The quantum mutual information $I(A:B)$ quantifies the total correlations between subsystems A and B and is computed directly as
\begin{equation}
    I(A:B) = S(\rho_A) + S(\rho_B) - S(\rho),
\end{equation}
where $S(\rho)$ is the von Neumann entropy, defined as
\begin{equation}
    S(\rho) = -\text{Tr}(\rho \log \rho)
    \label{Eq:VNE}
\end{equation}
\(\rho_A = \text{Tr}_B(\rho)\) is the reduced density matrix of subsystem A, and \(\rho_B = \text{Tr}_A(\rho)\) is that of subsystem B.

\begin{algorithm}
\caption{Classical Mutual Information}
\begin{algorithmic}[1]

\Function{CMI}{$\rho$}
    \State $\rho_A \gets \text{Tr}_B(\rho)$ 
    \State $\rho_B \gets \text{Tr}_A(\rho)$ 
    \State $f \gets 0$
    \For{$\theta \in [0, \pi], \phi \in [0, 2\pi]$}
        \State $\Pi_0 \gets \begin{pmatrix} \cos^2(\theta/2) & \frac{\sin\theta}{2} e^{-i\phi} \\ \frac{\sin\theta}{2} e^{i\phi} & \sin^2(\theta/2) \end{pmatrix}$
        \State $\Pi_1 \gets \begin{pmatrix} \sin^2(\theta/2) & -\frac{\sin\theta}{2} e^{-i\phi} \\ -\frac{\sin\theta}{2} e^{i\phi} & \cos^2(\theta/2) \end{pmatrix}$
        \State $H \gets 0$
        \For{$k \in \{0, 1\}$}
            \State $p_k \gets \text{Tr}((\Pi_k \otimes I) \rho (\Pi_k \otimes I)^\dagger)$
            \If{$p_k > 0$}
                \State $\rho_k \gets \text{Tr}_A\left( \frac{(\Pi_k \otimes I) \rho (\Pi_k \otimes I)^\dagger}{p_k} \right)$
                \State $H \gets H + p_k \cdot S(\rho_k)$
            \EndIf
        \EndFor
        \State $C \gets S(\rho_A) - H$
        \If{$C > f$}
            \State $f \gets C$
        \EndIf
    \EndFor
    \State $(\theta^*, \phi^*) \gets \arg \min_{\theta, \phi} f(\theta, \phi)$
    \State $J \gets -f(\theta^*, \phi^*)$
    \State \Return $J$
\EndFunction
\end{algorithmic}
\end{algorithm}

The classical mutual information $J(A:B)$, which measures the classical correlations accessible via local measurements, is computed using Algorithm 1. This algorithm optimizes over all possible projective measurements on subsystem A, parameterized by angles \(\theta \in [0, \pi]\) and \(\phi \in [0, 2\pi]\). For each pair \((\theta, \phi)\), it defines measurement operators
\begin{equation}
    \begin{split}
        \Pi_0 = \begin{pmatrix} \cos^2(\theta/2) & \frac{\sin\theta}{2} e^{-i\phi} \\ \frac{\sin\theta}{2} e^{i\phi} & \sin^2(\theta/2) \end{pmatrix} \\
        \Pi_1 = \begin{pmatrix} \sin^2(\theta/2) & -\frac{\sin\theta}{2} e^{-i\phi} \\ -\frac{\sin\theta}{2} e^{i\phi} & \cos^2(\theta/2) \end{pmatrix}
    \end{split}
\end{equation}

Then measurement probability $p_k$ and Conditional State $\rho_{B|k}$ at Bob's end $(k \in \{0,1\})$ are given by,
\begin{equation}
    p_k= \text{Tr}\left( (\Pi_k \otimes I) \rho \right), \rho_{B|k}= \text{Tr}_A \left( (\Pi_k \otimes I) \rho \right)/p_k
\end{equation}

Then we calculate classical mutual information using the following expression:
\begin{equation}
    J(A:B)= S(\rho_B) -  \min_{\theta, \phi} \sum_{k}p_kS(\rho_{B|k})
\end{equation}
Here $S(\rho)$ is the Von Neumann Entropy descibed in Eq.~\ref{Eq:VNE}. 

 \textbf{Entanglement of Formation:} The EoF is computed using Algorithms 2 and 3. Algorithm 2 calculates the concurrence, a standard measure of entanglement for two-qubit systems. Given a density matrix \(\rho\), it constructs the matrix \(R = \rho (Y \otimes Y) \rho^* (Y \otimes Y)\), where \(Y\) is the Pauli Y matrix and \(\rho^*\) is the complex conjugate of \(\rho\). It then computes the eigenvalues of \(R\), takes their square roots, sorts them in descending order as \(\lambda_1 \geq \lambda_2 \geq \lambda_3 \geq \lambda_4\), and defines the concurrence as
\begin{equation}
    C = \max(0, \lambda_1 - \lambda_2 - \lambda_3 - \lambda_4).
\end{equation}
Algorithm 3 uses this concurrence to determine the EoF. We compute the binary entropy
\begin{equation}
    h(x) = -x \log_2 x - (1 - x) \log_2 (1 - x),
\end{equation}
where $x = \frac{1 + \sqrt{1 - C^2}}{2}$, and $h(x)$ is returned as the EoF.

In summary, Algorithms 2 and 3 compute the EoF via concurrence and binary entropy, while Algorithm 1, combined with the direct calculation of \(I(A:B)\), enables the calculation of QD, providing a comprehensive assessment of entanglement and quantum correlations in the system.

 Finally, since we are calculating the Fidelity with respect to Bell states, we can modify Eq.~7, and simplify as, $F = \left( Tr \sqrt \sigma \right)^2$, where $\sigma$ is the noisy output.


\begin{algorithm}
\caption{Concurrence}
\begin{algorithmic}[1]
\Function{Concurrence}{$\rho$}
    \State $Y \gets \text{Pauli Y matrix}$
    \State $R \gets \rho \times (Y \otimes Y) \times \rho^* \times (Y \otimes Y)$
    \State $\text{eigenvalues} \gets \text{eigenvalues of } R$
    \State $\text{Eig} \gets \text{sort } \text{eigenvalues} \text{ in descending order}$
    \State $\lambda_1, \lambda_2, \lambda_3, \lambda_4 \gets \sqrt{\text{Eig}[0]}, \sqrt{\text{Eig}[1]}, \sqrt{\text{Eig}[2]}, \sqrt{\text{Eig}[3]}$
    \State $C \gets \max(0, \lambda_1 - \lambda_2 - \lambda_3 - \lambda_4)$
    \State \Return $C$
\EndFunction
\end{algorithmic}
\end{algorithm}

\begin{algorithm}
\caption{Entanglement of Formation}
\begin{algorithmic}[1]
\Function{EntanglementOfFormation}{$\rho$}
    \State $C \gets \text{Concurrence}(\rho)$
    \If{$C = 0$}
        \State \Return $0$
    \Else
        \State $x \gets (1 + \sqrt{1 - C^2}) / 2$
        \State $h \gets -x \log_2(x) - (1 - x) \log_2(1 - x)$
        \State \Return $h$
    \EndIf
\EndFunction
\end{algorithmic}
\end{algorithm}


\section{Quantum Correlations as Metrics}
\begin{figure} 
\centering
    \includegraphics[width=0.45\textwidth]{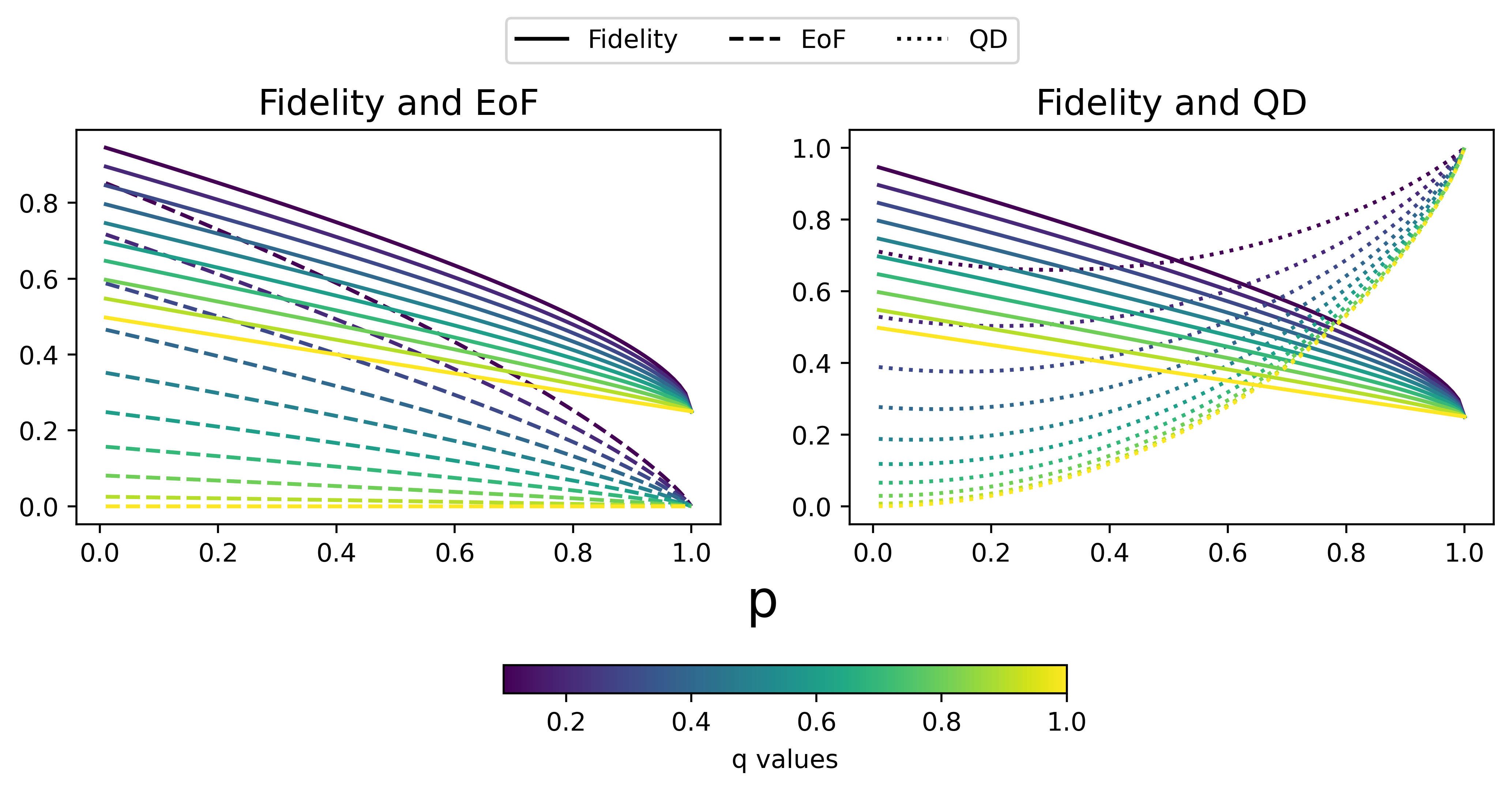}
    \captionof{figure}{Fidelity, QD and EoF. We have simulate for combined phase and amplitude damping. The x axis represents $p$, parameter for amplitude damping. Color represents $q$, parameter for phase damping. Fidelity, EoF and QD have been plotted for $(p,q)$ pairs. }
    \label{Fig:Fid}
\end{figure}

For our noise model, we calculate both \emph{Quantum Discord}~(QD) and \emph{Entanglement of Formation}~(EoF) to gauge how the channel degrades quantum correlations. As shown in Fig.~\ref{Fig:Fid}, We see when both QD and EoF are high, Fidelity is also high. And the decreased value of these metrics deteriorates fidelity. We also notice that QD can remain relatively high even when EoF is comparatively low, indicating that although the entanglement may be weak, there are still nonclassical correlations (captured by QD) available as a resource.

To further explore the relationship between these correlation measures and the effective quality of our quantum states, we perform a linear regression to determine whether a linear combination of QD and EoF can approximate fidelity. As summarized in Table~1, the model achieves a strong correlation ($R^2 \approx 0.98$ and $MSE < 10^{-3}$), with following regression model:
\begin{equation}
    \mathrm{Fidelity} \;=\; \alpha \;+\; \beta_{\mathrm{QD}} \,\mathrm{QD}
    \;+\;\beta_{\mathrm{EoF}} \,\mathrm{EoF}.
\end{equation}

These findings confirm that QD and EoF effectively encode the channel’s noise impact on our quantum states. Hence, they can serve as real-time indicators of whether the underlying entanglement is sufficiently robust.

\begin{table}[ht]
\centering
\begin{tabular}{l r}
\hline
\textbf{Terms} & \textbf{Value} \\
\hline
Mean Squared Error (MSE) & 0.000574 \\
$R^2$ Score       & 0.97787 \\
$\beta_{QD}$ & -0.23512 \\
$\beta_{EoF}$ & 0.76228 \\
$\alpha$ \ (Intercept)   & 0.47453 \\
\hline
\end{tabular}
\caption{Fidelity predicted from QD and EoF.}
\label{tab:linreg}
\vspace{-1em}
\end{table}

\section{Adaptive Purification}
The DEJMPS protocol effectively distills high-fidelity Bell states from noisy entangled pairs. However,
\begin{itemize}
    \item[1.] Each round sacrifices one entangled pair to purify another reducing the pair count, which is inefficient.
    \item[2.] DEJMPS relies on LOCC, which is suboptimal for co-located qubits where non-local operations are possible.
\end{itemize}

Hence we propose an adaptive protocol to tackle shortcomings with a design optimized for co-located qubits and adaptable to noise characteristics:
\begin{itemize}
    \item [1.] \textit{Preserving Pairs with Global Operations}: Joint operations across all qubits and ancillas aim to purify both pairs in a single round, avoiding DEJMPS’s pair loss.
    \item[2.] \textit{Flexible Error Handling}: Adjustable rotations enable tuning based on noise profiles, especially correlated errors, offering greater flexibility than DEJMPS’s rigid structure.

\end{itemize}

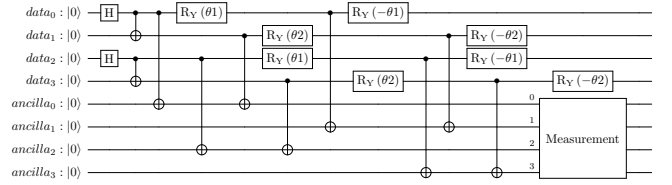
\begin{figure}
    \centering
    \scalebox{0.48}{
\Qcircuit @C=1.0em @R=0.2em @!R { \\
	 	\nghost{{data}_{0} : \ket{{0}} } & \lstick{{data}_{0} : \ket{{0}} } & \gate{\mathrm{H}} & \ctrl{1} & \ctrl{4} & \gate{\mathrm{R_Y}\,(\mathrm{{\ensuremath{\theta}}1})} & \qw & \qw & \ctrl{5} & \gate{\mathrm{R_Y}\,(\mathrm{-{\ensuremath{\theta}}1})} & \qw & \qw & \qw & \qw & \qw & \qw\\
	 	\nghost{{data}_{1} : \ket{{0}} } & \lstick{{data}_{1} : \ket{{0}} } & \qw & \targ & \qw & \qw & \ctrl{3} & \gate{\mathrm{R_Y}\,(\mathrm{{\ensuremath{\theta}}2})} & \qw & \qw & \qw & \ctrl{4} & \gate{\mathrm{R_Y}\,(\mathrm{-{\ensuremath{\theta}}2})} & \qw & \qw & \qw\\
	 	\nghost{{data}_{2} : \ket{{0}} } & \lstick{{data}_{2} : \ket{{0}} } & \gate{\mathrm{H}} & \ctrl{1} & \qw & \ctrl{4} & \qw & \gate{\mathrm{R_Y}\,(\mathrm{{\ensuremath{\theta}}1})} & \qw & \qw & \ctrl{5} & \qw & \gate{\mathrm{R_Y}\,(\mathrm{-{\ensuremath{\theta}}1})} & \qw & \qw & \qw\\
	 	\nghost{{data}_{3} : \ket{{0}} } & \lstick{{data}_{3} : \ket{{0}} } & \qw & \targ & \qw & \qw & \qw & \ctrl{3} & \qw & \gate{\mathrm{R_Y}\,(\mathrm{{\ensuremath{\theta}}2})} & \qw & \qw & \ctrl{4} & \gate{\mathrm{R_Y}\,(\mathrm{-{\ensuremath{\theta}}2})} & \qw & \qw\\
	 	\nghost{{ancilla}_{0} : \ket{{0}} } & \lstick{{ancilla}_{0} : \ket{{0}} } & \qw & \qw & \targ & \qw & \targ & \qw & \qw & \qw & \qw & \qw & \qw & \multigate{3}{\mathrm{Measurement}}_<<<{0} & \qw & \qw\\
	 	\nghost{{ancilla}_{1} : \ket{{0}} } & \lstick{{ancilla}_{1} : \ket{{0}} } & \qw & \qw & \qw & \qw & \qw & \qw & \targ & \qw & \qw & \targ & \qw & \ghost{\mathrm{Measurement}}_<<<{1} & \qw & \qw\\
	 	\nghost{{ancilla}_{2} : \ket{{0}} } & \lstick{{ancilla}_{2} : \ket{{0}} } & \qw & \qw & \qw & \targ & \qw & \targ & \qw & \qw & \qw & \qw & \qw & \ghost{\mathrm{Measurement}}_<<<{2} & \qw & \qw\\
	 	\nghost{{ancilla}_{3} : \ket{{0}} } & \lstick{{ancilla}_{3} : \ket{{0}} } & \qw & \qw & \qw & \qw & \qw & \qw & \qw & \qw & \targ & \qw & \targ & \ghost{\mathrm{Measurement}}_<<<{3} & \qw & \qw\\
\\ }}
    \caption{Adaptive Purification $U(\theta_1, \theta_2)$. We have used ancilla so that the subsystem have CPTP Maps. Following usual DEJMPS, we introduced rotation, but instead of $\theta$, we are using $\theta_1$ and $\theta_2$ since Alice and Bob's side had assymetric noise exposure.}
    \label{fig:adaptive}
\end{figure}

Fig. ~\ref{fig:adaptive} shows the circuit diagram for the adaptive purification protocol. Mathematically speaking, after decoding, the whole operation can be given by:
\begin{equation}
    \rho = \sum_k (\mathcal{E}_k \otimes I) (U_{ij} |\Phi^+\rangle \langle \Phi^+| ) (\mathcal{E}_k^\dagger \otimes I)
\end{equation}
Here $\mathcal{E}$ doesn't represent the quantum channel alone, it also captures the whole effect upto the decoding process. But we cannot use unitary map to reverse the process. So we use ancilla to make a CPTP Map.
After adding the ancilla $|0\rangle ^{\otimes r}$, and applying $U(\theta_1, \theta_2)$ to the whole system, then tracing out ancilla system $R$ can be written as:
\begin{equation}
    \rho'= Tr_{R}\left(U\left( (\rho\otimes (|0\rangle \langle 0|)^{\otimes r}) \right)U^\dagger\right) = \zeta(\rho) 
\end{equation}

We can theoretically design $U(\theta_1, \theta_2)$ such that, the map $\zeta$ is approximately the inverse of $\mathcal{E}\otimes I$, i.e.,
\begin{equation}
    \zeta \circ (\mathcal{E} \otimes I) \approx 1
\end{equation}
While inspired by DEJMPS, our protocol fundamentally differs, employing a global, ancilla-assisted unitary with adaptive controlled-rotations aiming for deterministic noise inversion.

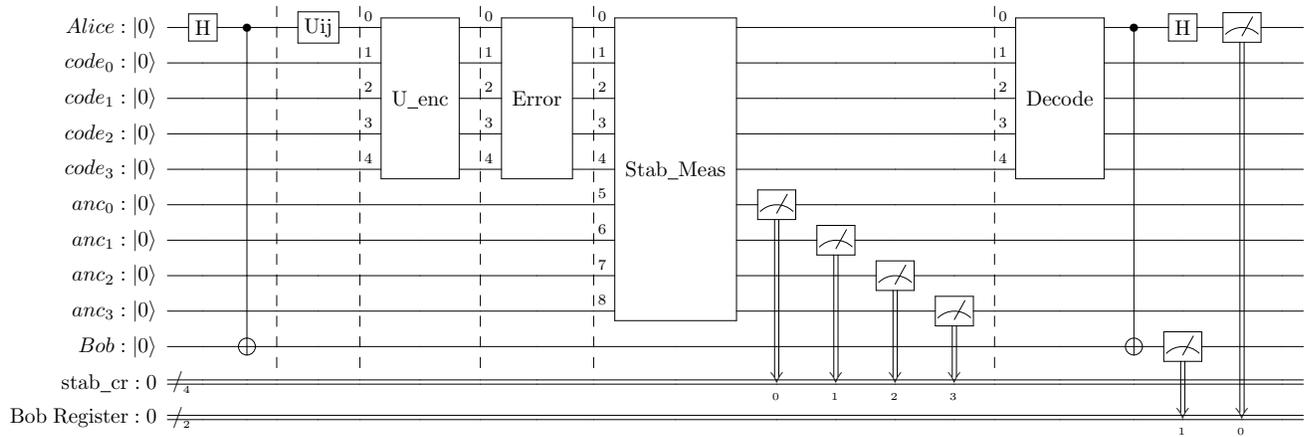
\begin{figure*}[!htb]
\centering
\scalebox{0.8}{
\Qcircuit @C=1.0em @R=0.2em @!R { \\
	 	\nghost{{Alice} : \ket{{0}} } & \lstick{{Alice} : \ket{{0}} } & \gate{\mathrm{H}} & \ctrl{9} \barrier[0em]{9} & \qw & \gate{\mathrm{Uij}} \barrier[0em]{9} & \qw & \multigate{4}{\mathrm{U\_{enc}}}_<<<{0} \barrier[0em]{9} & \qw & \multigate{4}{\mathrm{Error}}_<<<{0} \barrier[0em]{9} & \qw & \multigate{8}{\mathrm{Stab\_Meas}}_<<<{0} & \qw & \qw & \qw & \qw \barrier[0em]{9} & \qw & \multigate{4}{\mathrm{Decode}}_<<<{0} & \ctrl{9} & \gate{\mathrm{H}} & \meter & \qw & \qw\\
	 	\nghost{{code}_{0} : \ket{{0}} } & \lstick{{code}_{0} : \ket{{0}} } & \qw & \qw & \qw & \qw & \qw & \ghost{\mathrm{U\_{enc}}}_<<<{1} & \qw & \ghost{\mathrm{Error}}_<<<{1} & \qw & \ghost{\mathrm{Stab\_Meas}}_<<<{1} & \qw & \qw & \qw & \qw & \qw & \ghost{\mathrm{Decode}}_<<<{1} & \qw & \qw & \qw & \qw & \qw\\
	 	\nghost{{code}_{1} : \ket{{0}} } & \lstick{{code}_{1} : \ket{{0}} } & \qw & \qw & \qw & \qw & \qw & \ghost{\mathrm{U\_{enc}}}_<<<{2} & \qw & \ghost{\mathrm{Error}}_<<<{2} & \qw & \ghost{\mathrm{Stab\_Meas}}_<<<{2} & \qw & \qw & \qw & \qw & \qw & \ghost{\mathrm{Decode}}_<<<{2} & \qw & \qw & \qw & \qw & \qw\\
	 	\nghost{{code}_{2} : \ket{{0}} } & \lstick{{code}_{2} : \ket{{0}} } & \qw & \qw & \qw & \qw & \qw & \ghost{\mathrm{U\_{enc}}}_<<<{3} & \qw & \ghost{\mathrm{Error}}_<<<{3} & \qw & \ghost{\mathrm{Stab\_Meas}}_<<<{3} & \qw & \qw & \qw & \qw & \qw & \ghost{\mathrm{Decode}}_<<<{3} & \qw & \qw & \qw & \qw & \qw\\
	 	\nghost{{code}_{3} : \ket{{0}} } & \lstick{{code}_{3} : \ket{{0}} } & \qw & \qw & \qw & \qw & \qw & \ghost{\mathrm{U\_{enc}}}_<<<{4} & \qw & \ghost{\mathrm{Error}}_<<<{4} & \qw & \ghost{\mathrm{Stab\_Meas}}_<<<{4} & \qw & \qw & \qw & \qw & \qw & \ghost{\mathrm{Decode}}_<<<{4} & \qw & \qw & \qw & \qw & \qw\\
	 	\nghost{{anc}_{0} : \ket{{0}} } & \lstick{{anc}_{0} : \ket{{0}} } & \qw & \qw & \qw & \qw & \qw & \qw & \qw & \qw & \qw & \ghost{\mathrm{Stab\_Meas}}_<<<{5} & \meter & \qw & \qw & \qw & \qw & \qw & \qw & \qw & \qw & \qw & \qw\\
	 	\nghost{{anc}_{1} : \ket{{0}} } & \lstick{{anc}_{1} : \ket{{0}} } & \qw & \qw & \qw & \qw & \qw & \qw & \qw & \qw & \qw & \ghost{\mathrm{Stab\_Meas}}_<<<{6} & \qw & \meter & \qw & \qw & \qw & \qw & \qw & \qw & \qw & \qw & \qw\\
	 	\nghost{{anc}_{2} : \ket{{0}} } & \lstick{{anc}_{2} : \ket{{0}} } & \qw & \qw & \qw & \qw & \qw & \qw & \qw & \qw & \qw & \ghost{\mathrm{Stab\_Meas}}_<<<{7} & \qw & \qw & \meter & \qw & \qw & \qw & \qw & \qw & \qw & \qw & \qw\\
	 	\nghost{{anc}_{3} : \ket{{0}} } & \lstick{{anc}_{3} : \ket{{0}} } & \qw & \qw & \qw & \qw & \qw & \qw & \qw & \qw & \qw & \ghost{\mathrm{Stab\_Meas}}_<<<{8} & \qw & \qw & \qw & \meter & \qw & \qw & \qw & \qw & \qw & \qw & \qw\\
	 	\nghost{{Bob} : \ket{{0}} } & \lstick{{Bob} : \ket{{0}} } & \qw & \targ & \qw & \qw & \qw & \qw & \qw & \qw & \qw & \qw & \qw & \qw & \qw & \qw & \qw & \qw & \targ & \meter & \qw & \qw & \qw\\
	 	\nghost{\mathrm{{stab\_cr} : 0 }} & \lstick{\mathrm{{stab\_cr} : 0 }} & \lstick{/_{_{4}}} \cw & \cw & \cw & \cw & \cw & \cw & \cw & \cw & \cw & \cw & \dstick{_{_{\hspace{0.0em}0}}} \cw \ar @{<=} [-5,0] & \dstick{_{_{\hspace{0.0em}1}}} \cw \ar @{<=} [-4,0] & \dstick{_{_{\hspace{0.0em}2}}} \cw \ar @{<=} [-3,0] & \dstick{_{_{\hspace{0.0em}3}}} \cw \ar @{<=} [-2,0] & \cw & \cw & \cw & \cw & \cw & \cw & \cw\\
	 	\nghost{\mathrm{{Bob\;Register} : 0 }} & \lstick{\mathrm{{Bob\;Register} : 0 }} & \lstick{/_{_{2}}} \cw & \cw & \cw & \cw & \cw & \cw & \cw & \cw & \cw & \cw & \cw & \cw & \cw & \cw & \cw & \cw & \cw & \dstick{_{_{\hspace{0.0em}1}}} \cw \ar @{<=} [-2,0] & \dstick{_{_{\hspace{0.0em}0}}} \cw \ar @{<=} [-11,0] & \cw & \cw\\
\\ }}

\caption{Architecture Before Purification. Alice and Bob share an ebit, and Alice encodes her qubit. Then, she applies QEC encoding $U_{enc}$ and sends it through the quantum channel. Bob measures the stabilizer, performs correction operations accordingly, and decodes it.}
\label{fig:QCircuit}
\end{figure*}

\section{Proposed Architecture}
We assume that Alice and Bob initially share several pairs of maximally entangled qubits in the Bell state $\ket{\Phi^+}$. A subset of these pairs is used to transmit classical information, while another subset, referred to as \emph{pilot pairs}, is reserved for noise monitoring.
\begin{figure}[!htb]  
\centering
     \includegraphics[width=0.5\textwidth, alt={Proposed Architecture Diagram}]{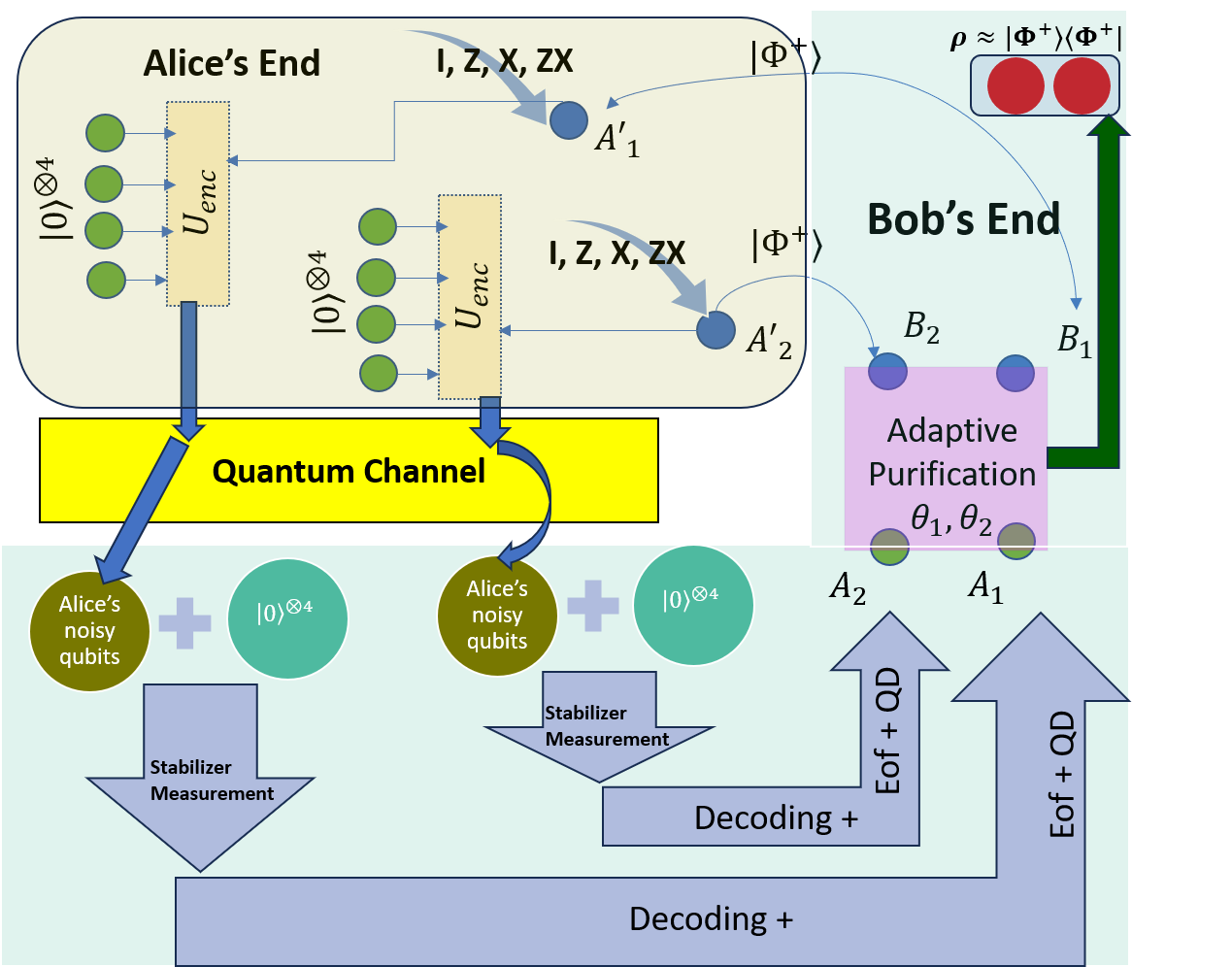}  
    \captionof{figure}{Proposed Architecture. Alice encodes her qubits and use 5 qubit perfect code. Then she sends them through channel, which Bob receives, decode, then applies adaptive purification using EoF, QD calculated from pilot pairs.}
    \label{fig:Architecture}
\end{figure}

\subsubsection*{1. Encoding and Transmission}:  
As shown in Fig.~\ref{fig:QCircuit} and Fig.~\ref{fig:Architecture}, Alice encodes two classical bits onto each data-carrying qubit using a local operation \(U_{ij}\). She then applies the five-qubit perfect code \(U_{\mathrm{enc}}\) to protect against single-qubit errors (e.g., amplitude and phase damping) before sending the qubits through a noisy channel.

\subsubsection*{2. Stabilizer Measurement and Decoding}:  
Bob receives the qubits and uses four ancillary qubits to perform stabilizer measurements, following the five-qubit code’s error correction process (Fig.~6). This recovers the logical state, though some noise may persist.

\subsubsection*{3. Pilot Pairs and Noise Estimation}:  
Pilot pairs, which pass through the same channel without carrying data, are measured for Quantum Discord (QD) and Entanglement of Formation (EoF). These metrics estimate noise levels without affecting the data pairs.

\subsubsection*{4. Adaptive Purification}:  
Using QD and EoF, Bob applies a global operation \(U(\theta_1, \theta_2)\) (Fig.~\ref{fig:adaptive}), tuned to counter \mbox{amplitude} and phase damping. Since Bob has all qubits locally, this preserves all entangled pairs and channel capacity.

\subsubsection*{5. High-Fidelity Outcome}:  
Bob recovers high-fidelity Bell pairs, enabling superdense coding or other quantum tasks. 

\section{Analysis and Discussion}
Our simulations and analytical discussion demonstrate that integrating the five-qubit perfect code with an adaptive, global purification approach significantly improves fidelity under damping noise channels. By monitoring quantum Discord (QD) and Entanglement of Formation (EoF) via pilot pairs, the protocol adjusts a global unitary to counter noise without discarding entangled pairs, unlike traditional LOCC-based purification, which sacrifices pairs and cuts channel capacity. Thus, our proposed architecture preserves all pairs and enhances resource efficiency for quantum communication.

Our findings, QD and EoF, strongly correlate with fidelity and capture distinct quantum correlation aspects. Their changes reveal noise type and magnitude, enabling adaptability crucial for unpredictable quantum networks.

Applying robust superdense coding with minimal overhead is invaluable for quantum networking, secure satellite communication, and advanced distributed quantum computing. By preserving all data‐carrying pairs, the method maintains higher effective throughput, enabling faster key distribution, more efficient classical‐bit transfer, and potentially improved multi‐user quantum communication. Using pilot pairs further allows the system to track time‐varying noise in extended networks, which is critical for large‐scale, real‐world implementations.

\section{Conclusion \& Future Work}
We have presented an adaptive superdense coding framework that integrates five‐qubit error correction with a global, non‐discarding purification strategy. By leveraging QD and EoF as indicators of noise severity, the protocol adjusts a global unitary to compensate for noise effect. Numerical simulations in amplitude and phase damping channels show promise in improvements of both fidelity and capacity over conventional fixed‐parameter or discard‐based methods. Immediate next steps involve numerically simulating the complete protocol to explicitly develop the mapping function from the measured correlations to the optimal purification parameters. Investigating the limits of its expressivity for inverting diverse noise channels also remains an avenue for future analysis. In addition, deep learning can be incorporated in the future to automate mapping correlations (QD and EoF) to optimal purification parameters as a scalable adaptive solution. In conclusion, our approach represents a key step toward achieving robust, high‐throughput quantum communication in realistic, dynamically changing environments.

\balance

\bibliographystyle{IEEEtran}

\bibliography{references}

\end{document}